\documentclass[12pt,preprint]{aastex}

\begin{document}

\title{Hydrogen Clouds before Reionization: a Lognormal Model Approach}

\author{Hongguang Bi\altaffilmark{1,2}, Li-Zhi Fang\altaffilmark{2}, 
Longlong Feng\altaffilmark{3,4} and Yipeng Jing\altaffilmark{5,6}}

\altaffiltext{1}{Shanghai Astronomical Observatory, Nandan Road 80, 
Shanghai 200030, China}
\altaffiltext{2}{Department of Physics, University of Arizona, Tucson, 
AZ 85721}
\altaffiltext{3}{Center for Astrophysics, University of Science
and Technology of China, Hefei, Anhui 230026, China}
\altaffiltext{4}{National Astronomical Observatories, Chinese
Academy of Science, Chao-Yang District, Beijing 100012, China}
\altaffiltext{5}{Shanghai Astronomical Observatory, the Partner Group 
of Max-Planck Institut f\"ur Astrophysik, Nandan Road 80,
Shanghai 200030, China}
\altaffiltext{6}{Beijing Astrophysical Center and Department of
Astronomy, Peking University, Beijing 100871, China}

\begin{abstract}

We study the baryonic gas clouds (the IGM) in the 
universe before the reionization with the lognormal (LN) model which is
shown to be dynamcially legitimate in describing the fluctuation
evolution in quasilinear as well as nonlinear regimes in recent years.
The probability distribution function (PDF) of the mass field in the 
LN model is long tailed and so plays an important role in rare events,
such as the formation of the first generation of baryonic objects.
Because in the this model the nonlinear field is directly mapped 
from the corresponding linear field, we can calculate density and velocity 
distributions of the IGM at very high spatial resolutions. We 
simulate the distributions at resolution of 0.15 kpc from $z=7$ to 15 
in the LCDM cosmological model. We performed a statistics of the 
hydrogen clouds at high redshifts, including column densities, clumping 
factors, sizes, masses, and spatial number density etc. One of our goals 
is to identify which hydrogen clouds are going to collapse. 
By inspecting the mass density profile and the velocity profile of clouds,
we found that the velocity outflow significantly postpones 
the collapsing process in less massive clouds, in spite of their masses 
are larger than the Jeans mass. That indicates that the formation of 
collapsed clouds with small mass at high redshift is substantial suppressed. 
Consequently, only massive ($>$ 10$^5$ M$_{\odot}$) clouds can form objects 
at higher redshift, and less massive (10$^4$-10$^5$ M$_{\odot}$) collapsed 
objects are formed later. Although the mass fraction in clouds 
with sizes larger than the Jeans length is already larger than 1\% at $z=15$, 
there is only a tiny fraction of mass ($10^{-8}$) in the clouds which are 
collapsed. If all the ionizing photons, and the $\sim 10^{-2}$ metallicity 
observed at low redshift are produced by the first 1\% mass of collapsed 
baryonic clouds, the majority 
of that first generation objects would be happen no much earlier than $z=10$.

\end{abstract}


\section{Introduction}

The lognormal (LN) model of the clustering of cosmic mass field was
used as a phenomenological description of the density and 
velocity distributions of the IGM (baryonic gas) in the redshift range 
$2 \leq z \leq 5$ 
in the study of the Ly$\alpha$ forest of QSO's absorption spectra
(Bi, 1993; Bi, Ge \& Fang 1995; Bi \& Davidsen 1997). Since then the 
LN model has gained substantial support from dynamical analysis
and numerical simulation of the nonlinear evolution of the cosmic mass field. 
First, it has been found that in the nonlinear regime the dynamics of the 
growth (irrotational) mode of density perturbations can be sketched by the 
random-force-driven Burgers' equation (Berera \& Fang 1994, Buchert, Dominguez 
\& Peres-Mercader 1999). An IGM model based on a random-force-driven 
Burgers' equation proposed by Jones (1999) yields intermittency of the 
IGM mass density field, and its probability distribution function (PDF) 
is found to be lognormal. Numerical simulations have 
directly shown that a lognormal PDF is a good approximation of the 
nonlinear density and velocity fields (e.g. Hui, Kofman \& Shandarin, 2000; 
Yang et al 2001). The LN model can be used to describe the cosmic 
gravitational clustering not only in quasilinear regimes, but also in highly 
nonlinear regimes. Recently, a detailed study on the dynamical system 
consisting of dark matters and the IGM supports also the LN model 
(Matarrese \& Mohayee 2002). 

In this paper, we will use the LN model to study semi-analytically the 
formation and evolution of the IGM  clouds before the 
reionization. A basic property of the LN model is that the PDF of the 
mass field is long tailed, and therefore it will play an important role 
in rare events, such as the formation of the first generation baryonic 
objects. A number of semi-analytic models have been developed to describe 
the structure formation in the early universe such as those based 
on extrapolations of the linear theory (e.g. Couchman \& Rees 1986, 
Madau, Meikin \& Rees 1997, Tozzi et al. 2000), analytic models (e.g. 
Tegmark et al. 1997, Miralda-Escude 1998, Valageas \& Silk 1999, 
Miralda-Escude, Haehnelt \& Rees 2000, Chiu \& Ostriker 2000),
the Press-Schechter (PS) formalism (e.g. Haiman \& Loeb 1998), and
numerical simulations (e.g. Abel et al. 1998, Bromm, Coppi \& Larson 1999,
Gnedin 2000,  Cen \& Haiman 2002). See also reviews in Barkana \& Loeb 
2001, and Madau 2002. As a semi-analytical approach, the non-linear field 
in the LN model are mapped from the linear density field. Therefore, one 
can take the advantage to simulate the spatial distributions of density 
and velocity, so as to calculate the density and velocity profiles of 
clumpy regions in the distribution, called hydrogen clouds thereafter. 

The Jeans length of the IGM is below 1 kpc before the reionization. To 
study hydrogen clouds on mass scale as small as the Jeans mass 
$\sim 10^4$ M$_{\odot}$, it is necessary to calculate the IGM distribution 
on scales at least as fine as 0.2 kpc. Using the LN model, we are able 
to simulate the IGM distribution on such small scales. We will focus on
the abundance of cloud going to collapse. Since the IGM cloud going to
collapse is a necessary condition of hosting star formation, the abundance 
of such collapsing clouds can set effective constrains on the formation of 
the first generation baryonic objects

The paper is organized as follows. In \S 2, we present an updated 
introduction of the LN model of the clustering of cosmic field. The basic 
features of the IGM clustering around the epoch of reionization will 
be discussed in \S 3. In \S 4, we analyze the statistic 
properties of the baryonic clumps (hydrogen clouds) based on the simulation 
of the LN model. In \S 5, we identify the clouds which match the
condition of going to collapse. Finally in \S 6, we summarize the results 
and our conclusions.

\section{Lognormal model updated}

The basic assumption of the LN model is that the PDF of the density
field $\rho({\bf x})$ is lognormal, i.e. 
\begin{equation}
\rho({\bf x}) = \bar{\rho_0}e^{X({\bf x})},
\end{equation}
where $X({\bf x})$ is a Gaussian field with mean $\overline{X}$ and 
variance $[\overline{ (X-\overline{X})^2]}^{1/2}=\sigma_0$. 
 From eq.(1), the mean of $\rho({\bf x})$ is 
\begin{equation}
\overline {\rho} = \bar{\rho_0}
 \exp\left [\overline{X}+
   \frac{1}{2}\overline{(X-\overline{X})^2}\right ].
\end{equation}
The overall average density $\overline {\rho}$ should always satisfy 
$\overline {\rho}=\bar{\rho_0}$, which is required by mass conservation. 
Therefore, eq.(2) requires
$\overline{X} = -(1/2)\overline{(X-\overline{X})^2}= -\sigma_0^2/2$. 
The density fluctuation or density contrast of the field is then 
$\delta =(\rho -1)/\bar{\rho}$.  

The PDF of the variable $\rho$ defined by eqs.(1) and (2) is
\begin{equation}
p(\rho/\bar{\rho})=\frac{1}{(\rho/\bar{\rho}) \sigma_0\sqrt{2\pi}}
 \exp\left [ -\frac{1}{2}
  \left(\frac{\ln (\rho/\bar{\rho}) + \sigma_0^2/2}{\sigma_0}\right )^2
   \right ], \hspace{3mm} \rho \geq 0.
\end{equation}
Obviously, it is normalized as $\int_{0}^{\infty}p(x)dx =1$. The 
variance of $\rho$  is given by (Vanmarcke 1983)
\begin{equation}
\sigma=[e^{\sigma_0^2}-1]^{1/2}.
\end{equation}
It means the second moment to be
$\overline {(\rho/\bar{\rho})^2}=\exp({\sigma_0^2})$. The high order 
moments is 
\begin{equation}
\overline {(\rho/\bar{\rho})^n}=
   \exp \left [(n^2-n)\frac{\sigma_0^2}{2}\right ].
\end{equation}
Equation (5) shows 
$[\overline {(\rho/\bar{\rho}^n}]^{1/n} >
[\overline {(\rho/\bar{\rho})^2}]^{1/2}$, i.e.
the moment is divergent when $n \rightarrow \infty$. This indicates that 
the PDF eq.(3) is long tailed. The probability of 
long tail events $\rho \gg 1$ given by eq.(3) is much larger than that 
of a Gaussian PDF. 

Now we should identify the Gaussian field $X({\bf x})$. When the density
perturbation is small, $\delta ({\bf x}) \ll 1$, the mass field should be in
linear regime $\delta_0({\bf x})$, i.e. we need $\delta({\bf x}) \simeq 
\delta_0({\bf x})$. This yields $X=\delta_0({\bf x}) -\sigma_0^2/2$, or
\begin{equation} 
\rho({\bf x}) = \bar{\rho}_0\exp[\delta_0 ({\bf x}) -  \sigma_0^2/2],
\end{equation}
where $\sigma_0=\langle \delta_0 ^2 \rangle ^{1/2}$ is the variance of the
linear Gaussian field of $\delta_0 ({\bf x})$ on the scale $R$ considered.
When $\sigma_0$ is small, eq.(4) gives $\sigma = \sigma_0$. Equation (6) 
is a basic assumption of the LN model. With eq.(6), the nonlinear mass field, 
$\rho({\bf x})$, is given by an exponential mapping from the corresponding 
linear density field $\rho_0({\bf x})$ or $\delta_0({\bf x})$. 

The first successful application of the LN model mapping eq.(6) is to model 
the IGM distribution observed by the QSO Ly-$\alpha$ forests (Bi 1993 and 
Bi \& Davidsen 1997). However, the LN model is not only useful in weakly 
nonlinear regime, but also for highly nonlinear evolution. The dynamical 
study of the lognormal PDF can be traced back to the adhesion approach, 
which sketches the nonlinear evolution of the growth mode of the cosmic 
gravitational clustering with the Burgers' equation (Gurbatov, Saichev \& 
Shandarin 1989). This equation yields reasonable mass function of clumps
(Vergassola et al. 1994). Considering the stochastic nature of field variables, 
the cosmic clustering actually should be described by the random-force-driven 
Burgers' equation (Berera \& Fang 1994; Buchert, Dominguez \& Peres-Mercader 
1999). On the other hand, the lognormal PDF is found to be good approximation 
of density field described by this equations. Therefore, the log-normal model 
is dynamically legitimate to describe the cosmic clustering in quasilinear as 
well as nonlinear regimes. This is also supported by numerical simulation.
The one-point distribution of the cosmic mass and velocity fields on nonlinear 
regime, such as the number density of galaxies, pairwise velocity, angular 
momentum, etc from observed data and simulation samples, are in good agreement 
with lognormal distribution (e.g. Kofman et al. 1994, Hui, Kofman, \& Shandarin 
2000, Yang et al 2001, Pando, et al 2002). 

\section{IGM distribution with lognormal model}

\subsection{Jeans length of baryonic matter}

In a homogeneous universe with the mean mass density $\bar{\rho}$, the 
Jean length of gaseous baryonic matter or the IGM is defined by
$\lambda_b\equiv v_s(\pi/G\bar{\rho})^{1/2}$, where $v_s$ is the sound 
speed of the gas. For the current study, it is convenient to
use a comoving scale $x_b$ given by 
\begin{equation}
x_b \equiv \frac{1}{H_0}\left[\frac{2\gamma k_BT_m}
{3\mu m_p\Omega(1+z)}\right]^{1/2},
\end{equation}
which is 2$\pi$ times smaller than $\lambda_b$. In eq.(7), $T_m$ and 
$\mu$ are the mean temperature and molecular weight of the 
gas, $\Omega$ is the cosmological density parameter of total mass and 
$\gamma$ the ratio of specific heats. The corresponding Jeans mass is
$m_J=(4\pi/3)\lambda_b^3\bar{\rho}$.

Primordial baryons, created at the time of nucleosynthesis, recombines 
with electrons to become neutral gas at $z \sim 1000$, which corresponds 
to the Jeans mass of about $10^6$  $M_{\odot}$. 
Before $z=137$, the residual ionization of the cosmic gas keeps its
temperature locked to the CMB temperature. After that the gas cools 
down adiabatically during expanding of the universe. Assuming $\gamma=5/3$,
i.e. hydrogen temperature $T\propto \rho_b^{2/3}$, where $\rho_b$ is
the mean mass density of baryonic matter, the evolution of the comoving 
Jeans length will depend approximately on $(1+z)$. At $z=10$, the hydrogen 
temperature is $\sim 1.8$ K (Medvigy \& Loeb 1991), the Jeans length 
$x_b\sim 1$ kpc, and the Jeans mass drops to about $ 10^4$ 
$M_{\odot}$. Note that the Jeans mass is 6 magnitudes smaller than 
that of galaxies we observed. Figure 1 plots the comoving Jeans 
length $x_b$ as a function of the cosmic scale factor $1/(1+z)$. 

During the reionization, the baryonic gas will be heated by UV ionizing 
background with temperature from $1.8$ K to $\sim 1.3 \times 10^4$ K;
and there is a 2 magnitudes increasing in $x_b$ or 6 magnitudes in the 
Jeans mass (e.g. Ostriker \& Gnedin 1996). In Fig. 1, the reionization
is assumed to happen at $z=7$ instantly, i.e. the IGM is assumed
to be almost completely neutral and ionized before and after 
the reionization, respectively. It should be emphasized that the 
assumption of reionization redshift $z=7$ does not affect on the 
result of clustering {\it before} reionization discussed below. That is, 
for instance, the calculation of clustering at $z=12$ is independent   
of whether the reionization redshift is at 7 or 10. 

After the reionization, the IGM temperature is maintained at about 
$\sim 10^4$ K by the UV background photons, and therefore, $x_b$ will 
gradually increase with the decrease of $z$ due to the factor $(1+z)$ 
in eq.(1). Note that the HeII ionization occurs at redshift 3.3 that leads 
to an increase of the IGM temperature from $1.3\times 10^4$ K before 
$z=3.3$ to $2.5\times 10^4$ K after $z=3.3$ (Theuns et al 2002). This is 
shown in Fig. 1 by the small jump of $x_b$ at the redshift $\sim 3.3$. 

\subsection{IGM mass density field in lognormal model}

The linear evolution of the IGM mass field, $\rho_0({\bf x})$, driven by 
the gravity of dark matter mass field, $\rho_{dm}({\bf x})$, has been 
studied by using various assumptions of the thermal property of the IGM (e.g.
Nusser 2000, Matarrese \& Mohayee 2002). A 
common conclusion is 
\begin{equation}
\delta_0({\bf k},t) = (1+ {\rm decaying \ terms})\delta_{dm}({\bf k},t) 
  + {\rm decaying \ terms}
  \hspace{1cm} {\rm if} \ k \ll k_J.
\end{equation}
where $\delta_0({\bf k},t)$ and $\delta_{dm}({\bf k},t)$ are, respectively, 
the Fourier transform of density fluctuations of the IGM 
$\delta_0({\bf x})= [\rho_0({\bf x})/\overline{\rho}_0]-1$ and the dark matter,
$\delta _{dm} ({\bf x})= [\rho_{dm}({\bf x})/\overline{\rho}_{dm}]-1$.
The decaying terms in Eq.(10) depend on the initial conditions. 
That is, regardless 
specific assumptions of thermal processes and initial condition, the linear 
fluctuations of the IGM mass field on scales larger than the Jeans 
length $x_b$ always follows the dark matter mass field (Fang et al 1993)
\begin{equation}
\delta _0({\bf x}) = \frac{1}{4\pi x_b ^2} \int 
\frac{\delta _{dm} ({\bf x}_1)}
{|{\bf x} - {\bf x}_1|}e^{-\frac{|{\bf x} - {\bf x}_1|}{x_b}} d{\bf x}_1, 
\end{equation}
Obviously, $\langle\delta_0 \rangle =\langle \delta _{dm}\rangle=0$.

With the LN model of \S 2, the nonlinear mass field
of the IGM, $\rho_b({\bf x})$, has to be is given by an exponential mapping 
from the corresponding linear density field of dark matter $\rho_0({\bf x})$ 
as
\begin{equation}
\rho_b({\bf x}) = \bar{\rho}_b \exp[ \delta_0 ({\bf x}) - 
  \sigma_0^2/2 ],
\end{equation}
where $\sigma_0$ is the variance of the linear Gaussian field of 
$\delta_0 ({\bf x})$ on the scale of the Jeans length. To simplify 
eq.(12), we use normalized the mean baryonic matter density 
$\bar{\rho}_b=1$ below. 

The number of  $\sigma_0$ is shown in Fig. 2, in which, we use the 
low density flat cold dark matter model (LCDM) with the density parameter 
$\Omega_0=0.3$, the cosmological constant $\Omega_{\Lambda}=0.7$ and
the Hubble constant $h=0.7$. The linear power spectrum $P(k)$ is 
given by the fitting formula given by Eisenstein \& Hu (1998). The 
linearly increasing of $\sigma_0$ with cosmic factor $1/(1+z)$ is due 
to the linear increasing of the density perturbations. The variance 
$\sigma$ of the LN PDF [eq.(3)] is also plotted in Fig. 2.
We see from Fig. 2 that the first time of the variance $\sigma_0$ 
reaching to order of one is in the period of redshift $15 > z > 7$. 
This should be the epoch of the first generation collapsed object 
formation.

 When the fluctuations $\delta_0({\bf x})$ are small, 
$\delta_0 ({\bf x}) \ll 1$, eq.(10) yields 
$\delta({\bf x}) \simeq \delta_0({\bf x})$. It is the linear solution 
eq.(8). On the other hand, on scales $x \leq x_b $, eq.(10) gives
the nonlinear relation between the IGM density distribution and 
the dark matter gravitational potential as
\begin{equation}
\rho_b({\bf x}) \propto \exp \left[-\frac{\mu m_p}{\gamma k T} 
\psi_{dm}({\bf x}) \right ].
\end{equation}
Eq.(11) is the well-known isothermal hydrostatic solution, which describes 
highly clumped structures such as intracluster gas (Sarazin \& Bahcall 1977). 
Therefore, the density distribution $\rho_b({\bf x})$ in eq.(10) is 
consistent with highly nonlinear distribution of the baryonic mass in the 
dark matter gravitational potential wells. Note that the primordial potential 
$\psi_{dm}$ remain linear and 
Gaussian much longer time than the density (Brainerd, Scherrer \& 
Villumsen 1983; Bagla \& Padmanabhan 1994), eq.(11) directly shows that the 
IGM field $\rho_b({\bf x})$ is lognormal. Thus, if we require that the 
mapping between the IGM field $\rho_b({\bf x})$ and the linear dark matter 
field $\delta_{dm}({\bf x})$ should satisfy 1.) the linear relation eq.(8) 
in linear regime; 2.) the exponential relation eq.(11) in highly nonlinear 
regime, the mapping of eqs.(10) probably is the most reasonable one.  
Therefore, the lognormal mapping eq.(10) can uniformally describe the IGM 
distribution from linear, weakly nonlinear, to highly nonlinear regimes, 
corresponding to density contrast $\ll 1$, $\simeq 1$, and $\gg 1$ 
respectively, without introducing extra parameters before the dynamics of 
clouds are dominated by hydro processes, such as cooling and heating during 
star formation.  
 
The LN model eq.(9) - (10) is successful to fit the transmitted
flux of the QSO Ly$\alpha$ absorption spectrum (Bi \& Davidsen 1997; 
Feng \& Fang, 2001). The lognormal PDF of the IGM field also gives better
fitting to the recent detected intermittency of the transmitted flux with 
high resolution, high signal to noise ratio samples of QSO Ly$\alpha$ 
absorption spectra (Jamkhedkar, Zhan, \& Fang, 2000, Zhan, Jamkhedkar, 
\& Fang, 2001, Feng, Pando, \& Fang 2001, Feng, Pando \& Fang 2003, 
Jamkhedkar et al. 2003). 

\subsection{Volume filling factor and cumulative mass fraction of 
   IGM clouds}

We now study the formation and collapsing of overdense hydrogen clouds 
in the IGM field with the lognormal mapping. First we demonstrate the  
lognormal PDF with the volume filling factor $V(>\rho)$, which is 
the fractions of volume with density larger than a given $\rho$. From 
eqs.(3) or (10) we have 
\begin{equation}
V(>\rho) = \int ^{\infty}_{\rho} p(\rho) d\rho
= \frac{1}{2} {\rm erfc}\left (\frac{\sigma_0}{2\sqrt{2}} + 
 \frac{\ln \rho}
{\sqrt{2} \sigma_0} \right ).\
\end{equation}
Figure 3 shows $V(>\rho)$ for $\sigma_0$ on the Jeans length scales 
at redshifts $z=30$, 20, 10 and 7. For a uniform Gaussian random field, 
roughly we have half space volume with density lower than the mean 
$\rho =1 $, and half larger than $\rho =1 $. However, Fig. 3 shows 
that $V(>1) < 1/2$ even when $z=30$. That is, the mass field of the IGM 
has already deviated from a Gaussian field at very early time. Fig. 3 shows
also that most IGM has fallen into clumps and most space is occupied 
by very low density gas. This is why the lognormal model is able to 
fit intermittent fields (Pando, et al. 2002; Feng, Pando \& Fang, 2003). 
In these fields, the mass is concentrated in peaks or spikes, which 
are randomly and widely scattered in space with a low mass density 
surrounded. 

The long tail of the LN model can more easily be seen with the 
cumulative mass fraction $M(>\rho)$, which is the fraction of mass in 
regions having mass density larger than a given $\rho$, given by 
\begin{equation}
M[>(\rho/\bar{\rho})] = \int ^{\infty}_{\rho/\bar{\rho}} x p(x) dx
= \frac{1}{2} {\rm erfc}
\left (\frac{\ln (\rho/\bar{\rho})}{\sqrt{2} \sigma_0}-
\frac{\sigma_0}{2\sqrt{2}}\right ). 
\end{equation}
Figure 3 shows $M(>\rho)$ for $\sigma_0$ on the Jeans length scales at 
redshifts $z=30$, 20, 10 and 7. The curves of $M(>\rho)$ at high $\rho$ 
shows clearly the long tail. For instance, for the curve of $z=7$, the 
number of $M(>100)$ is less $M(>10)$ only by a factor of about 10. That 
is, the mass fraction of large mass events ($\rho=100$) in $R$ can be 
10\% of small mass events ($\rho=10$). This is because that the 
variance $\sigma_0$ at $z=7$ is about 1.8. Objects of $\rho=10$  
correspond to $\ln \rho /1.8 \simeq 1.3$ $\sigma_0$ event, and 
$\rho=100$ to $\ln \rho /1.8\simeq 2.6$ $\sigma_0$ event. Therefore, the 
probabilities for  $\rho=10$ and $\rho=100$ are different only by a 
factor about 10. On the other hand, if the PDF is Gaussian,
$\rho=10$ objects corresponds to 1.3-$\sigma_0$ event, while high density 
$\rho=100$ corresponds to 7.7-$\sigma_0$ event. In this case, the number 
of $\rho=100$ objects is completely negligible with respect to 
$\rho=10$ objects. 

Figure 3 shows that $M(>\rho)$ is significant for $\rho =10$ and $z > 10$. 
On the other hand, we have always $V(>\rho) \sim 0$ with 
$\rho =10$, regardless redshift. This indicates that a significant part 
of mass concentrates in a small volume. That is dense objects at high 
redshifts. The long tail events may not be important at low redshift, 
as at that time, $\sigma_0$ on large scales is close to 1, and the 
formation of large mass objects is no longer to be rare events. 

\section{Statistics of Hydrogen Clouds}

\subsection{Simulations of the IGM distribution}

To study the clustering and collapsing of hydrogen clouds, we produce
simulation samples of spatial distribution of gas $\rho({\bf x})$ 
with the LN model developed in last section. In order
to quickly grasp the features of these distributions, we will simulate 
1-D distribution. The details of the simulation procedure has been given in 
Bi \& Davidsen (1997). A brief description is as follows. 

We first simulate the 1-D density and velocity distribution in the 
Fourier space, $\delta_0(k)$ and $v(k)$, which are two Gaussian 
random fields. Both $\delta_0(k)$ and $v(k)$ are given by the 
power spectrum, $P_0(k)$, as follows (Bi 1993; Bi, Ge \& Fang 1995) 
\begin{equation}
\delta _0 (k, z) = D(z) (u(k) + w(k)),
\end{equation}
\begin{equation}
v(k, z) = F(z) \frac{H_0}{c} ik \alpha (k) w(k),
\end{equation}
where $D(z)$ and $F(z)$ are the linear growth factors for fields 
$\delta _0( x)$ and $v(x)$ at redshift $z$. The fields $w(k)$ and $u(k)$
are Gaussian with power spectra given by 
\begin{equation}
P_w(k) = \alpha ^{-1} \int ^\infty _k P_{0}(q) 2\pi q^{-1} dq,
\end{equation}
\begin{equation}
P_u(k) = \int ^\infty _k P_{0}(q) 2\pi q dq - P_w(k),
\end{equation}
where $P_0(k)$ is the power spectrum of the 3-D field $\delta_0({\bf x})$.
Functions $\alpha(k)$ in eq.(16) is defined by
\begin{equation}
\alpha(k) = \frac{\int ^\infty _k P_{0}(q) q^{-3} dq}
            {\int ^\infty _k P_{0}(q) q^{-1} dq}.
\end{equation}
 From eq.(9), we have
\begin{equation}
P_{0}(k) = \frac{P_{dm}(k)}{(1+x_b ^2 k^2)^2}
\end{equation}
where $P_{dm}(k)$ is the dark matter power spectrum in 3-D.
Thus, for a given $P_{dm}(k)$ and $x_b$, one can produce the distributions 
$\delta_0(k)$ at grid points $k_i$, $i=1, 2, ..., N$ in the 
Fourier space. The spatial distributions in the real line-of-sight 
space $\delta_0(x)$ can be obtained by using a Fast Fourier Transform.
Since the velocity follows the linear evolution longer than the density, we 
can use the linear $v(k)$ and its Fourier counterpart as velocity field.  

The Jeans length before the reionization is about one h$^{-1}$kpc. It 
requires the resolution of simulation to be less than 0.2 kpc. Our 
simulation range is 40 Mpc in comoving space. The total number of pixels 
is 262144, so the pixel size is 0.152588 kpc. Three random samples of the 
density field with 1 Mpc comoving size are plotted for $z=15$, $10$ and 
$7$ in Fig. 4. The prominent spikes correspond to the dense hydrogen 
clouds which are candidates of the collapsing objects. Such hydrogen clouds 
occur not only at redshift $z=7$, but also at $z=15$. Although the number and 
height of the spikes decrease with higher redshifts, this 
$z$-dependence actually is not very sharp. The typical size of the clouds 
shown in Fig. 4 is of the Jeans length, but they have different height. This 
indicates that the probability of the spike events on the same size is not 
sharply dependent on the their height, i.e. the probability does not sharply 
depend on the mass density $\rho$ of the spikes, but only on $\ln \rho$. This 
is an effect of the lognormal long tail.

\subsection{Basic properties of hydrogen clouds}

To identify hydrogen clouds, we first smooth the simulated density field 
with a Gaussian 
filter of dispersion $x_b$. With the smoothed field, baryonic or hydrogen
clouds are identified as the regions between two successive minima in 
the field. For each clouds, we have a mass density profile between the 
two minima. The maximum between the two successive minima is the central 
position of the cloud. The width $D$ is defined to be the FWHM of the top 
density. The column density $N_{HI}$ of an absorber is obtained by 
summing gas densities in each pixel from the first minimum to the next 
minimum in the smoothed field. For each cloud, we 
assign a peculiar velocity to be the velocity at the top density, and 
the internal velocity profile is the peculiar velocity of each pixel 
relative to the center velocity.

Since column density depends on both hydrogen number density and the size
of clouds, it may not be a good indicator for the density contrast of 
hydrogen clouds. For instance, two clouds with the same column density 
may have 10 times difference in their density contrast due to a 10 times 
difference of their sizes. If such cases are common, one can not measure 
mass density with the column density. However, statistically we have a good 
reason to use column density $N_{HI}$ to characterize the mass density of 
baryonic clouds. Fig. 5 plots the relation between the column density $N_{HI}$ 
and density $\rho$ of clouds identified from one realization of the 1-D field. 
Fig. 5 shows a tight correlation between the column density $N_{HI}$ and the 
mass density $\rho$ of clouds. For a given $N_{HI}$, the dispersion 
of $\rho$ is 
no larger than 20\%. That is, the column density is mainly determined by the 
cloud mass density, not their size. A similar correlation has also been 
found among the pre-collapsed halos identified from N-body simulation samples 
(Xu, Fang \& Wu 2000). Thus, the IGM clustering in dark age can be 
approximately described by the statistics of the hydrogen clouds with 
the number $N_{HI}$.

We first calculate $N_c(>N_{HI}, z)$ h Mpc$^{-1}$, which is the 1-D 
comoving number density of clouds with column densities larger than a 
given $N_{HI}$. The redshifts are taken to be $z=7$, 10 and 15. 
The differential number density is $dN_c(>N_{HI}, z)/d\ln(N_{HI})$, 
which is plotted in Figure 6a. We can see from Fig. 6a that clouds 
with $N_{HI} \geq 10^{20}$ at $z=7$ are much more than that at $z=15$. 
It indicates that most clouds $N_{HI} \geq 10^{20}$ formed at redshifts 
less than 15.

Figure 6b presents the cumulated mass fraction of clouds with column 
density larger than a given $N_{HI}$. It also shows that mass fraction of 
clouds with $N_{HI}>10^{20.5}$ underwent a significant evolution from 
$z=15$ to 7. The mass fraction of $N_{HI}>10^{20}$ clouds at $z=15$ 
is about 1\%, but it is about 10\% at $z = 7$. For clouds with  column 
densities $10^{19.0}< N_{HI}< 10^{19.75}$, the number density and mass 
fraction at redshifts 7 to 15 are comparable. 

Figures 6c and 6d give, respectively, the mean top density and the mean 
comoving width $D$ of $N_{HI}$ clouds. Figures 6c and 6d show that the 
clouds with $N_{HI} \geq 10^{20}$ at $z=7$ have smaller size (width) 
and higher central density than that at $z=15$ and 10. This shows that 
hydrogen clouds in collapsing in the period from $z=15$ to $z=7$. The mass 
density of $N_{HI} \geq 10^{19.25}$ clouds has overdensity $\rho > 5$ at all 
redshifts. It means that the $N_{HI} \geq 10^{19.25}$ clouds were already 
going to collapse as early as $z=15$.

To study the 3-D statistics of the clouds, we assume that most clouds are 
spherical in the 3-D space with scale $D$. This is equal to approximate 
the non-linear dynamical evolution of a clouds as a spherical collapsing
process. This approximation probably is poor for clouds with low $N_{HI}$, 
but would be better for higher $N_{HI}$. Thus, we can estimate the baryonic 
mass of the cloud by $\rho m_p D^3$. The result is plotted in Figure 7a. 
We see that the $M$-$N_{HI}$ relation is insensitive to redshift.  
Fig. 7b is the cumulated mass fraction of 3-D clouds with mass larger 
than a given $M$. It is interested to see that the mass fraction of clouds 
with mass in the range $M < 10^{4.5}$ M$_{\odot}$ is weakly dependent on 
redshift from $z=7$ to 15. The mass fraction for $M \geq 10^{4.5}$ M$_{\odot}$ 
is significantly dependent on redshift. 

Figure 8a is the 3-D cumulative number density of clouds with column density 
larger than a given $N_{HI}$. Figure 8b is similar to Figure 8a, but it is          
with respect to the mass $M$ of clouds. We see again that the cumulative 
number densities of clouds at $N_{HI} \simeq 10^{19.25}$ or 
$M \simeq 10^{4.5}$ M$_{\odot}$ are not a strong function of redshifts from 
7 to 15. The cumulative number densities of massive clumps $N_{HI} > 10^{20}$ 
or $M > 10^5$ M$_{\odot}$ decreases with higher redshift quickly.
 From these statistics, we can conclusion that, in the LN model,
the baryonic clouds with mass $M > 10^{5}$ M$_{\odot}$ underwent a strong 
evolution in the epoch of $15 > z > 7$. But for clouds with mass 
$M < 10^{4.5}$ M$_{\odot}$, the evolution is moderate.

\section{Clouds going to collapsing}
   
\subsection{Density profile and collapsing}

We now identify which clouds are going to collapse. Obviously collapsed 
IGM clouds is a necessary condition for hosting star formation. 
To study the details of the collapsing, we will calculate the density 
profile within each cloud. The mean densities profile of clouds 
along the line-of-sight have been plotted in Fig. 9, in which the 
horizontal axis is the comoving distance from the center of the clouds, 
and redshifts are taken to be 15, 12, 10 and 7. For each redshift, the 
successive curves from lower to higher correspond to column density 
$\log N_{HI}= 16.25+m\times 0.5$, and $m=0,1,...8$.   
 
 Fig. 9 shows that all density profiles are approximately
exponential, i.e. $\rho \propto \exp(-x/2 x_0)$, where $x_0$ is 
of the order of $x_b$. The exponential profile comes from the 
exponential factor in eq.(11). 
In Fig. 9, we can see that the comoving density profile of clouds with 
$N_{HI}=10^{17.75}$ is almost independent of redshift $z=15$ to 7.
That is, the comoving size of clouds of $N_{HI}=10^{17.75}$ is almost 
independent of redshift, and therefore, they are expanding with the 
Hubble flow. These clouds are comoving with the Hubble expansion.
This is expected. From Fig. 5, we seen that $N_{HI}=10^{17.75}$ corresponds 
to about $\rho =1$, i.e. these ``clouds'' have the same density
as background universe. It should follow the Hubble expansion.

For clouds with $N_{HI}<10^{17.75}$, or $\rho <1$, the comoving size 
is bigger with smaller redshift. In other words, these ``clouds'' 
actually are in voids. They are a part of the voids. Therefore, their
expansion is faster than Hubble streaming.

For clouds with $N_{HI} \geq 10^{18.25}$ the comoving size is 
smaller for smaller redshifts. This is, there are in the phase of turn 
around. The mean density of these clouds at $z=7$ is $\rho \sim 6$ 
(Fig. 6). These clouds have decoupled from the Hubble expansion. Their 
expansion is slower than the Hubble expansion. However, we should not
simply identify all clouds with $N_{HI}\geq 10^{18.25}$ to be in 
collapsing, because they may still be in physical expanding, but only 
the expanding velocity is less than the Hubble velocity.
   
\subsection{Velocity profile and collapsing}

Whether a cloud is collapsing should also be investigated through its 
velocity profile. We calculate the peculiar velocity profile 
for all clouds identified from the 3,000 realizations of 1-D IGM field. 
The mean 1-D velocity profiles for various $N_{HI}$ clouds are plotted in 
Fig. 10, in which the horizontal axis is the comoving distance from 
the center of cloud, and velocity is measured with respect to the 
center of clouds. The redshifts are taken to be 15, 12, 10 and 7. 
For each redshift, the successive curves from smaller to larger 
velocity correspond to column density 
$\log N_{HI}= 16.25+m\times 0.5$, and $m=0,1,...8$. 

Figure 10 also shows the Hubble flow $v_H$ with respect to the center 
of the clouds. The dotted lines of Fig. 10 actually is $-v_H$. 
The physical velocity profiles is then given by $v+v_H = v - (-v_H)$. 
Fig. 11 is the same as Fig. 10, but shows a zoom-in profile of the central 
parts of the clouds. 

In Figs. 10 and 11, we can classify baryonic clouds into three types. 
1.) velocity profile of $v$ is negative, $v<0$, on the side of 
separation $<0$ and positive, $v>0$, on the side of separation $>0$. 
These clouds are expanding away from their centers, i.e. their physical 
expansion velocity is faster than the Hubble flow. 2.) the velocity profile 
of $v$ is positive, $v>0$, on the side of separation $<0$ and negative, 
$v<0$, on the side of separation $>0$, but the physical velocity profile is 
negative, $v+v_H <0$, on the side of separation $<0$ and positive, 
$v+v_H >0$, on the side of separation $>0$. These clouds are decoupled 
from the Hubble flow, but they still remain in expanding phase, and not 
reached yet to their maximum expansion.  3.) the physical velocity profile is 
positive, $v+v_H \geq 0$, on the side of separation $<0$ and negative, 
$v+v_H \leq 0$, on the side of separation $>0$. These clouds are in the 
phase of maximum expansion, or starting to collapse. In other words, its 
infall velocity is greater than the Hubble velocity in physical space.

Obviously, only clouds of type (3) can be identified as collapsing or 
collapsed clouds. That is, in Figs. 10 and 11, only the $N_{HI}$ clouds 
having the $v$ velocity profiles (solid line) equal to or higher than the 
negative Hubble flow $-v_H$ (dotted line) are the possible spots of star 
formation. Therefore, not all clouds with size larger than the Jeans length 
are in the collapsing phase. Although clouds with small $N_{HI}$ have size 
larger than the Jeans length, the velocity outflow of clouds significantly 
postpones their collapsing. This leads to a suppression of formation of low 
mass collapsed clouds in high redshifts. This result is consistent with
the simulation result (Gnedin, 2000), which shows that the formation of 
low mass objects in the redshift range $12 > z > 7$ is significantly 
suppressed with respect to the Jeans length prediction. 

\subsection{Constrains on the first star formation}

Besides the Jeans length of IGM in eq.(9), we did not introduced any other
parameters related to the hydro processes of star formation in this paper. 
Even so, we can already set some constrains on the formation of first 
generation stars by considering that collapsed clouds are the necessary 
environment of hosting star formation. 

As mentioned in last subsection, only clouds of type (3) are collapsed or
collapsing, and can play the role of the host of star formation. Thus, the
first generation IGM clustering can be seen from the properties of the type 
(3) clouds, and their redshift-dependence. Table 1 lists the basic properties 
of these clouds, including the minimum column densities $(\log N_{HI})_{min}$, 
the minimum mass $M_{min}$ (M$_{\odot}$) and the minimum density $\rho_{min}$.
All the minimum density $\rho_{min}$ of the collapsing clouds are $>10$ or 
even $>>10$. Note $\sigma_0$ is of order of one, the events of 
$\rho_{min} >10$ are really rare, or very rare. Therefore, the first 
generation of star formation rely on the long tail of the PDF.

\begin{table}[t]
\caption{Properties of collapsing clouds vs. redshift $z$}
\bigskip
\begin{tabular}{lllll}
\tableline
 redshift & 15 & 12 & 10 & 7  \\
\tableline
 $(\log N_{HI})_{min}$ & $22.25$  & $20.25$ & $19.5$ & $18.75$ \\
 $M_{min}$ (M$_{\odot}$) & $10^{6.5}$  & $10^{5.3}$  & $10^{4.4}$  
    & $10^{3.5}$ \\
 $\rho_{min}$    & $3.1\times 10^3$  & $1.9\times 10^2$  & 35 & 19 \\
\tableline
\end{tabular}
\end{table}

Figure 12 shows the redshift dependence of the cumulative mass fraction 
$M(>M_J)$ of clouds larger than Jeans mass, and $M(>\rho_{min})$ of clouds 
in collapsing and collapsed baryonic objects. One can clearly see from
Fig. 12 that the collapsing of hydrogen clouds is significantly suppressed
when $z >12$. The evolution of $M(>M_J)$ is moderate in the redshift range of 
$7< z < 15$, while the redshift-dependence of $M(>\rho_{min})$ is dramatic
when $z>12$. This is because the suppression of collapsing is stronger at 
higher redshift, and weaker at lower redshift. At early universe $z=15$, 
there is already about 1 \% mass in the clouds with size larger than the 
Jeans length. However, there is only a tiny fraction of mass ($\sim 10^{-8}$) 
in the clouds which are collapsed. This is because at redshift 15 only 
clouds with mass $> 10^{6.5}$ M$_{\odot}$ can collapse, and hosting star 
formation. The mass fraction in the collapsed clouds is higher for smaller 
redshift. Therefore, the formation history of the first-generation baryonic 
objects lasts the entire epoch from $z =15$ to $10$. 

It has been estimated that if all the ionizing photons, and the 
$\sim 10^{-2}$ metallicity observed at low redshift are produced by the 
first generation stars, the mass fraction of the IGM falling into 
the collapsed clouds should not be much less than 1\% 
(Haiman \& Loeb 1997, Ostriker \& Gnedin 1996) before the reionization. 
Using the 1\% mass to quantify the first generation of baryonic 
objects, we can conclude from Table 1 and Fig. 12 that the generation of 
source responsible for reionization is likely to occur not much earlier 
than z=10, and therefore, the reionization itself should take place not 
much earlier than that era.

\section{Conclusion and discussion}

We have outlined an evolution picture of the IGM clustering with the 
LN model, which is capable to sketch cosmic clustering of both
the dark matter and the IGM in weakly nonlinear, as well as nonlinear 
regimes. With the fine numerical simulations, we can make a number of 
important statistics of the IGM clouds before the 
reionization. We focus, in particular, the abundance of hydrogen clouds 
going to collapse, which would be the candidates for the harbor of the first 
generation stars. While the detail of the first star formation 
depends on many chemical and hydro factors, the formation of the 
collapsing candidates is a necessary condition to form stars or 
galaxies. Therefore, we can set up effective constraint on the history 
of the formation of the first objects. 

A remarkable feature of the LN model is that the PDF of mass field
is long tailed. The long tail events, such as the formation of massive 
clouds at high redshifts, have a much larger probability than a Gaussian 
field. The rare events of the first generation collapsed clouds are 
sensitive to the long tail. For a Gaussian PDF, there are very few events 
of mass perturbation much larger than $\sigma_0$, and therefore, there 
is no massive objects formed at early time for a Gaussian PDF. However,
the LN model predicted a much high probability of forming massive 
clouds at high redshift.

The LN model provides the density as well as the velocity distributions of the 
IGM. Comparing the density profile with velocity profile of clouds, one can 
reveal that clouds with the Jeans mass are not collapsing immediately 
at high redshift. Typically, the IGM cloud around a group of dark matter 
potential wells will stay in the phase of slow expanding in proper space 
for a while. The smaller the mass of hydrogen cloud, the longer the 
time to remain in this phase. Only clouds with high enough mass can 
enter the phase of collapsing in proper space. Thus, as shown in Table 1,
the minimal mass of hydrogen clouds in collapsing is larger for higher 
redshifts. That is, the clustering in the LN model is not of the bottom-up 
hierarchy at scales less than $10^{5-6}$ M$_{\odot}$. Contrarily, massive 
collapsed clouds formed  in the first, and less massive clouds later. It 
should be pointed out that this feature cannot simply be explained with the 
redshift-dependence of the Jeans mass. Although the Jeans mass is smaller 
at smaller redshift, it is not enough to explain the significant evolution 
of $M_{min}$ from $10^{6.5}$ to $10^{4.4}$ M$_{\odot}$ in the redshift 
range $15 < z<10$. This feature is again because the suppression of the 
formation of collapsed clouds with small mass on high redshifts. The lognormal 
mapping makes the evolution of $M_{min}$ to be much stronger than that of the 
Jeans mass $M_J$.

It should be point out that there are factors leading to overestimate the
collapsed hydrogen clouds. First, we count all clouds in the type 3 (\S 5.2)
to be the host of star formation. Obviously, the collapsing clouds in the 
type 3 are not yet to be that host. Second, the total mass of baryons in 
the collapsed objects is estimated by the mass of collapsed clouds. This may
also be a source of the overestimation. The last but not least, the LN models 
seems to overestimate the abundance of saturated absorption lines of 
Ly$\alpha$ (Matarrese \& Mohayaee 2002), it may indicate the excess of dense 
clouds. Nevertheless, all these factors actually are to strengthen the 
result on the upper limit to the mass fraction in the first generation 
objects at a given redshift, and then, to consolidate the conclusion
that the generation of source responsible for reionization and the 
reionization itself is likely to occur not much earlier than z=10.  

We may improve the overestimation problem by considering
the hydrodynamics of the IGM. However, we would rather like to trade the
overestimation with the uncertainty of IGM hydrodynamics. That is
in the current approach, the conclusions are not dependent on parameters 
related to the hydro processes of star formation such as cooling time scale, 
rate of star formation, and efficient of producing reionization photons by 
stars etc. Our conclusions are, however, sensitively dependent on the power 
spectrum of initial density perturbations on few h$^{-1}$ kpc. Therefore,
it would be useful to get information of the initial power spectrum on
such small scales.   

\acknowledgments

We thank Dr. Ping He and Mr. Hu Zhan for helps during the preparation
of this paper. We thank also the referee and editor for their helpful 
comments.

\newpage

\figcaption {The Jeans length of baryonic gas (IGM) as a function of 
the cosmic factor $1/(1+z)$.
} \label{Fig1}

\figcaption {The variances of the density fluctuations at the Jeans length 
as a function of the cosmic scale factor. The dot line is for the linear 
fluctuation, the solid line is for the non-linear evolution given by 
the LN model, i.e. $\sigma^2=\exp (\sigma_0 ^2) -1$.
 } \label{Fig2}

\figcaption {Volume filling factor (top) and cumulative mass (bottom) as 
function of the scale factor $a = (1+z)^{-1}$ at redshifts $z=20$, 12, 10 
and 7 in the CDM model.
} \label{Fig3}

\figcaption {Three samples within 1 Mpc size randomly chosen 
from $z=15$ (top), $z=10$(middle) and $z=7$(bottom) simulation. 
The X-axis is the line-of-sight distance in the comoving space
and the Y-axis is the overdensity.
 } \label{Fig4}

\figcaption {Relation between $\rho$ and column density $N_{HI}$ of hydrogen
clouds identified from one realization 
} \label{Fig5}

\figcaption {Statistics of hydrogen clouds as function of its column density 
$N_{HI}$. (a) 1-D differential number density (h Mpc$^{-1}$)
$dN_c(>N_{HI})/d\ln N_{HI}$; 
(b) mass fraction of clouds with column density $>N_{HI}$; (c) mean 
top density $\rho$ and (d) width (h$^{-1}$ kpc). The redshifts are taken 
to be $z=7$(solid), 10(dotted) and 15(dashed).
} \label{Fig6}

\figcaption {(a) Hydrogen mass of clouds as a function of column density;
(b) cumulated mass friction of clouds with mass $>M$. The redshifts are 
taken to be $z=7$(solid), 10(dotted) and 15(dashed).} \label
{Fig7}

\figcaption {3-D number density of clouds as functions of (a) column density 
$N_{HI}$, and (b) mass $M$ of clouds. The redshifts are taken to be 
$z=7$(solid), 10(dotted) and 15(dashed). 
} \label{Fig8}

\figcaption {Density profiles of clouds at redshifts 15, 12, 10 and 7. 
For each redshift, the successive curves from lower to 
higher correspond to column density $\log N_{HI}= 16.25+m\times 0.5$, 
and $m=0,1,...8$.} \label{Fig9}

\figcaption {Velocity profiles of clouds at redshifts 15, 12, 10 and 7. 
For each redshift, the successive curves from lower to 
higher correspond to column density $\log N_{HI}= 16.25+m\times 0.5$, 
and $m=0,1,...8$.
} \label{Fig10}

\figcaption {The same as Fig. 11, but for central parts of clouds.
} \label{Fig11}

\figcaption {The cumulative mass fraction of $M(>M_J)$ and $M(>\rho_{min})$
as function of redshift.
} \label{Fig12}


\begin{references}

\reference{} Abel, T., Anninos, P., Norman, M.L. \& Zhang, Y. 1998, 
   \apj, 508, 518, 

\reference{} Abel, T., Bryan, G. \& Norman M. 2000, \apj, 540, 39

\reference{} Bagla, J.S., \& Padmanabhan, T. 1994, \mnras, 266, 227

\reference{} Barkana, R., Loeb, A, 2001, astro-ph/0010468

\reference{} Berera, A. \& Fang, L.Z. 1994, Phys. Rev. Lett., 72, 458

\reference{} Bi, H.G. 1993, \apj, 405, 479 
 
\reference{} Bi, H.G \& Davidsen, A. F. 1997, \apj, 479, 523. 

\reference{} Bi, H.G., Ge, J., \& Fang, L.Z. 1995, \apj, 452, 90 

\reference{} Brainerd, T.G., Scherrer, R.J., \& Villumsen, J.V. 1993, 
\apj, 418, 570

\reference{} Bromm, V. Coppi, P.S. \& Larson, R. B. 1999, \apj, 527, L5

\reference{} Buchert, T., Dominguez, A. \& Peres-Mercader, J.,  1999,
  A\&A, 349, 343

\reference{} Cen, R. \& Haiman, Z. 2002, \apj, 542, L75

\reference{} Couchman, H.M.P. \& Rees, M.J., 1986, \mnras, 221, 53

\reference{} Chiu, W.A. \& Ostriker, J. 2000, \apj, 534, 507

\reference{} Eisenstein, D. \& Hu, W. 1998, \apj, 496, 605

\reference{} Fang, L.Z., Bi, H.G., Xiang, S.P., \& B\"{o}rner, G. 
   1993, \apj, 413, 477 

\reference{} Feng L.L. \& Fang, L.Z. 2001, \apj, 535, 519

\reference{} Feng, L.L. Pando, J. \& Fang, L.Z. 2003, \apj, 587, 487

\reference{} Gnedin, N. Y. 2000, \apj, 542, 535

\reference{} Gurbatov, S.N., Saichev, A.I. \& Shandarin, S.F. 1989,
  \mnras, 236, 385

\reference{} Hui, L., Kofman, L. \& Shandarin, S. F., 2000, \apj, 537, 12

\reference{} Jamkhedkar, P., Zhan, H. \& Fang, L.Z. 2000, \apj, 543, L1

\reference{} Jamkhedkar, P.,  Feng, L.L., Zheng, W., Kirkman, D., 
      Tytler, D.  \& Fang, L.Z. 2003, \mnras, in press

\reference{} Jones, B.T., 1999, \mnras, 307, 376

\reference{} Kofman, L., Bertschinger, E., Gelb, J.M., Nusser, A., 
   \& Dekel, A. 1994, \apj, 420, 44

\reference{} Haiman, Z. \& Loeb, A. 1998, \apj, 503, 505

\reference{} Madau, P., 2002, astro-ph/0212555

\reference{} Madau, P., Meikin, A. \& Rees, M. J. 1997, \apj, 475, 429

\reference{} Matarrese, S. \& Mohayaee, R., 2002, \mnras, 329, 37

\reference{} Medvigy, D. \& Loeb, A. 2001, astro-ph/0110014

\reference{} Miralda-Escude, J. 1998, \apj, 501, 15, 

\reference{} Miralda-Escude, J., Haehnelt, M. \& Rees, M.J., 2000, 
   \apj, 530, 1 

\reference{} Nusser, A., 2000, \mnras, 317, 902. 

\reference{} Ostriker, J.P. \& Gnedin, N.Y. 1996, \apj, 472, L63

\reference{} Pando, J.,  Feng, L.L., Jamkhedkar, P., 
  Zheng, W., Kirkman, D., Tytler, D.  \& Fang, L.Z. 2002, \apj, 574, 575

\reference{} Peebles, P.J.E. 1993, Principles of Physical Cosmology 
(Princeton: Princeton Univ. Press)

\reference{} Press \& Schechter 1974, \apj, 187, 425

\reference{} Shraiman B.I. \& Siggia, E.D. 2000, Nature, 405, 8

\reference{} Tegmark, M., Silk, J., Rees, M.J., Blanchard, A., Abel, T., 
\& Palla, F. 1997, \apj, 474, 1, 

\reference{} Theuns, T., Schaye, J., Zaroubi, S., Kim, T.-S., Tzanavaris, P.,
 \& Carswell, B. 2002, \apj, 567, L103

\reference{} Tozzi, P., Madau, P., Meiksin, A., \& Rees, M. J. 2000, 
   \apj, 528, 59, 

\reference{} Valageas, P. \& Silk, J. 1999, A\&A, 347, 1

\reference{} Vergassola, M., Dubrulle, B., Frisch, U. \& Noullez, A. 1994,
    A\&A, 289, 325. 
 
\reference{} Vanmarcke, E., 1983, {\it Random Field}, (MIT, Mass)

\reference{} Xu, W., Fang, L.Z. \&  Wu, X.P. 2000, \apj, 532, 728

\reference{} Yang, X.H., Feng, L.L., Chu, Y.Q. \&  
     Fang, L.Z. 2001, \apj, 560, 549.

\reference{} Zhan, H., Jamkhedkar, P. \& Fang,L.Z. \apj 555, 58

\end{references}
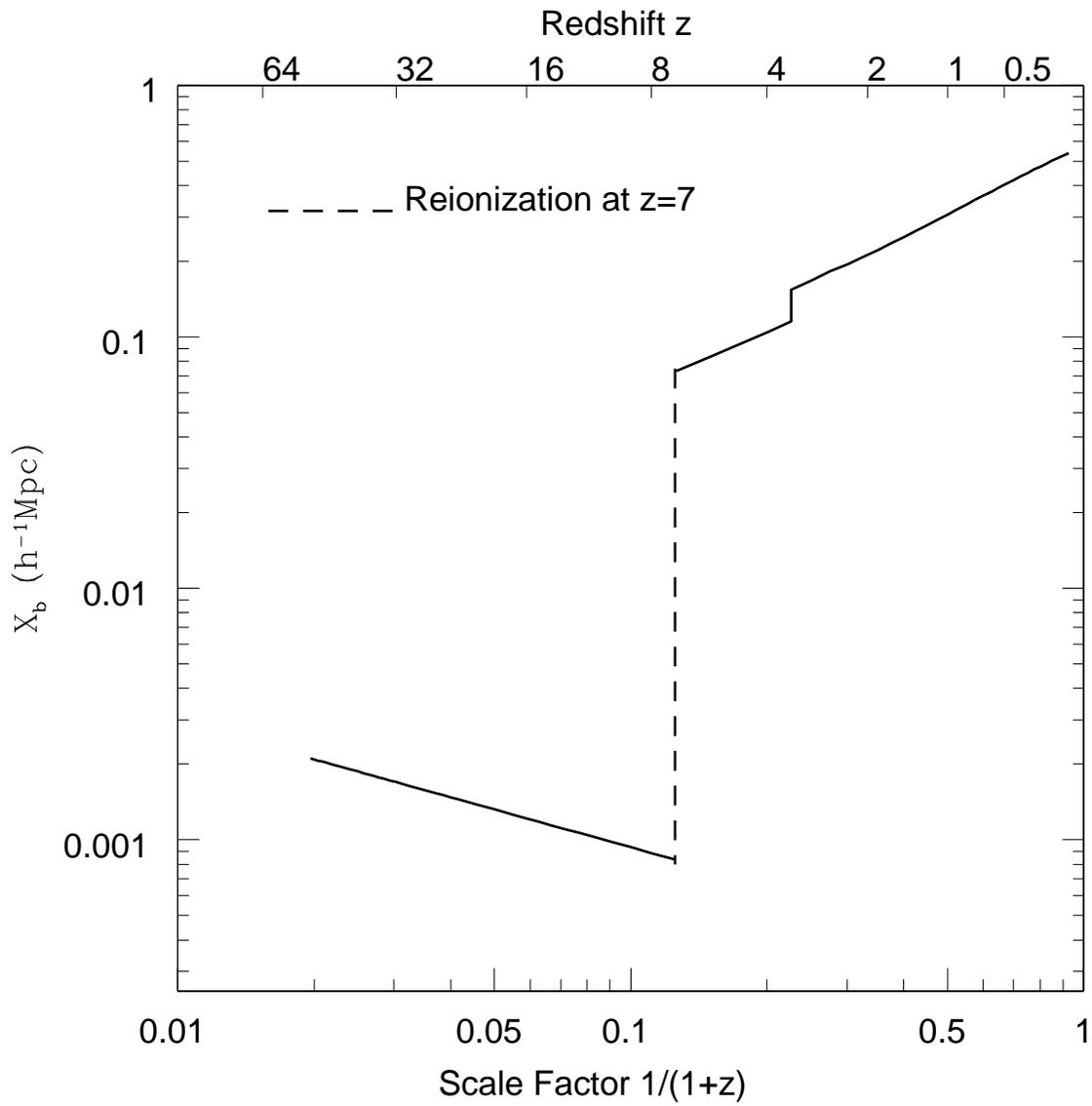
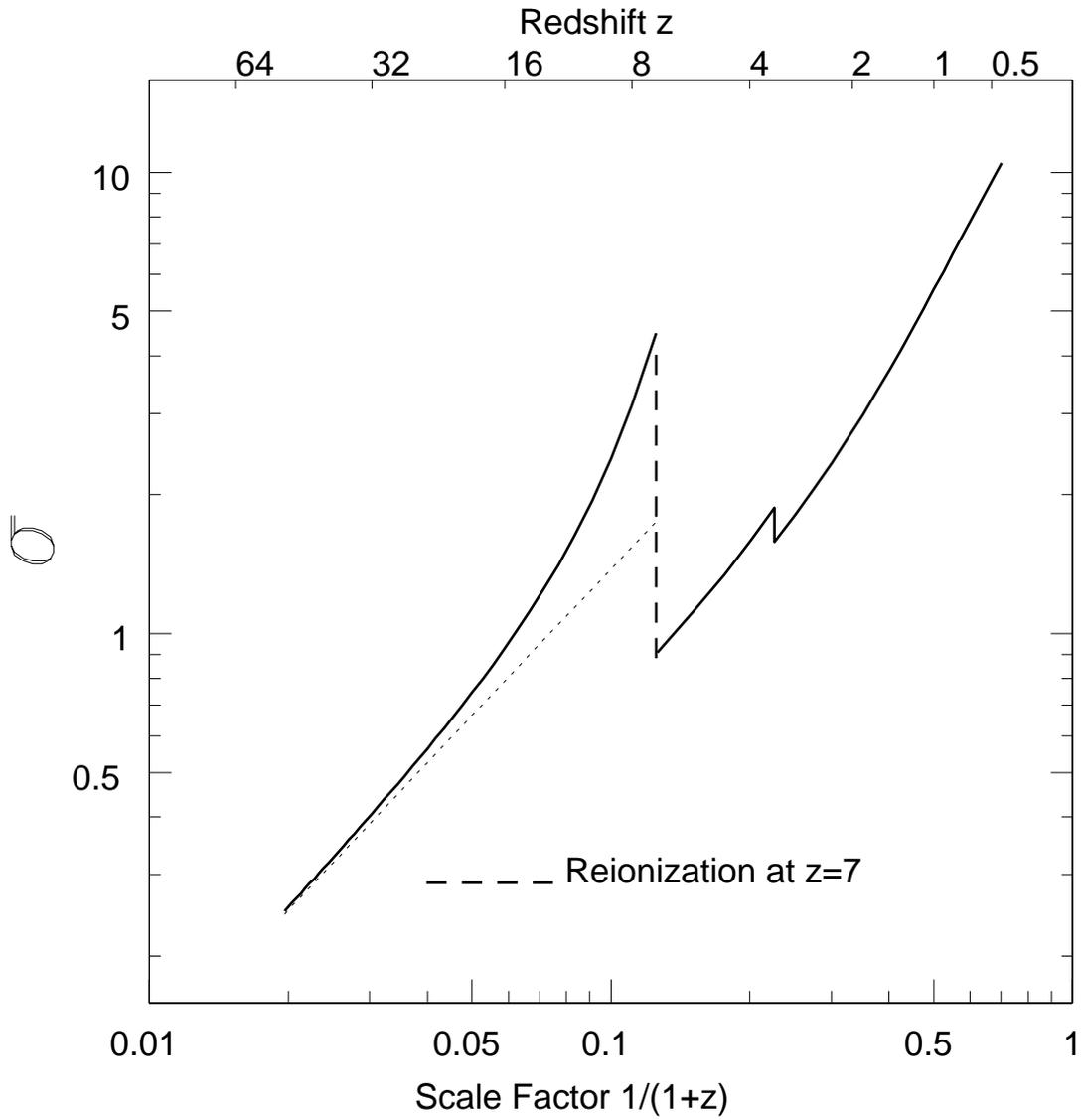
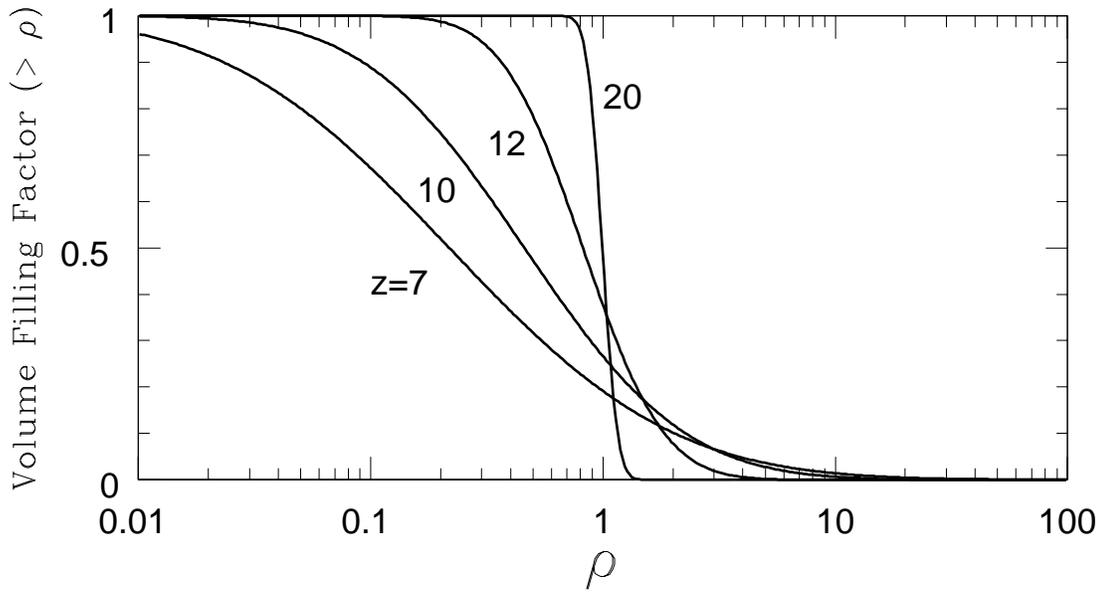
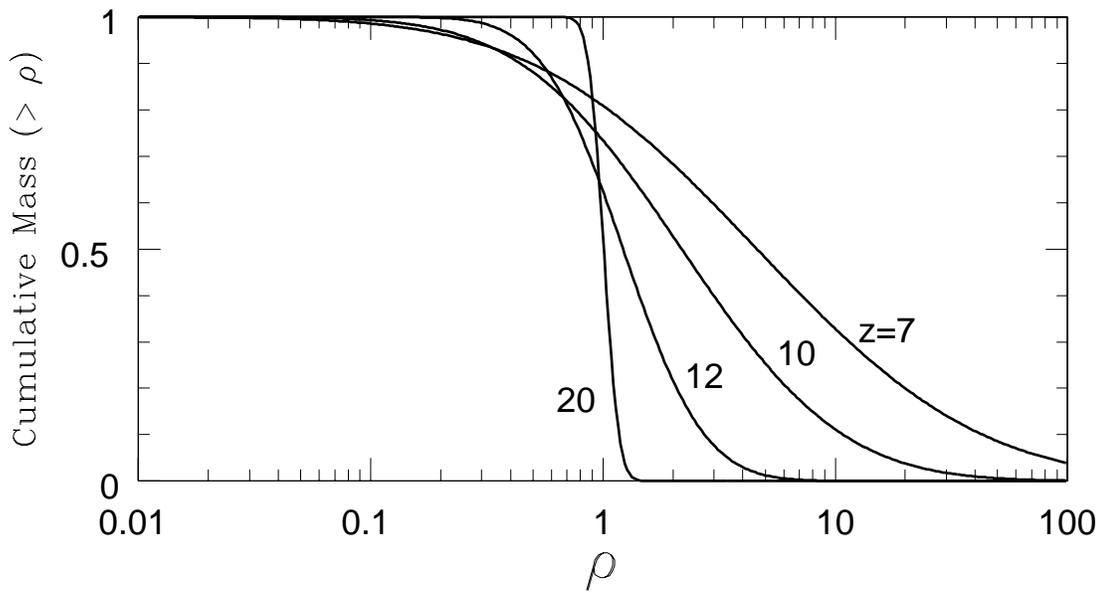
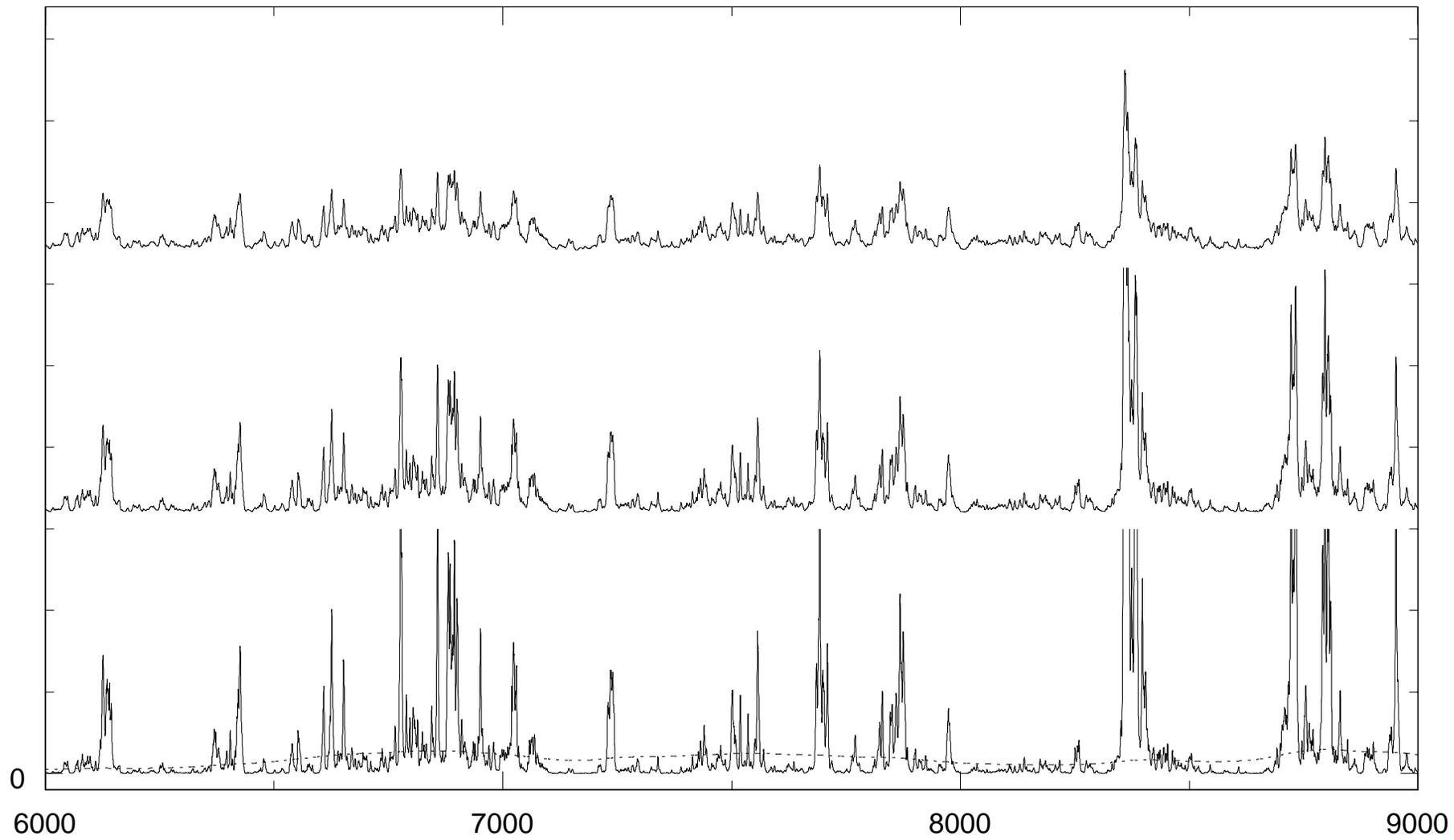
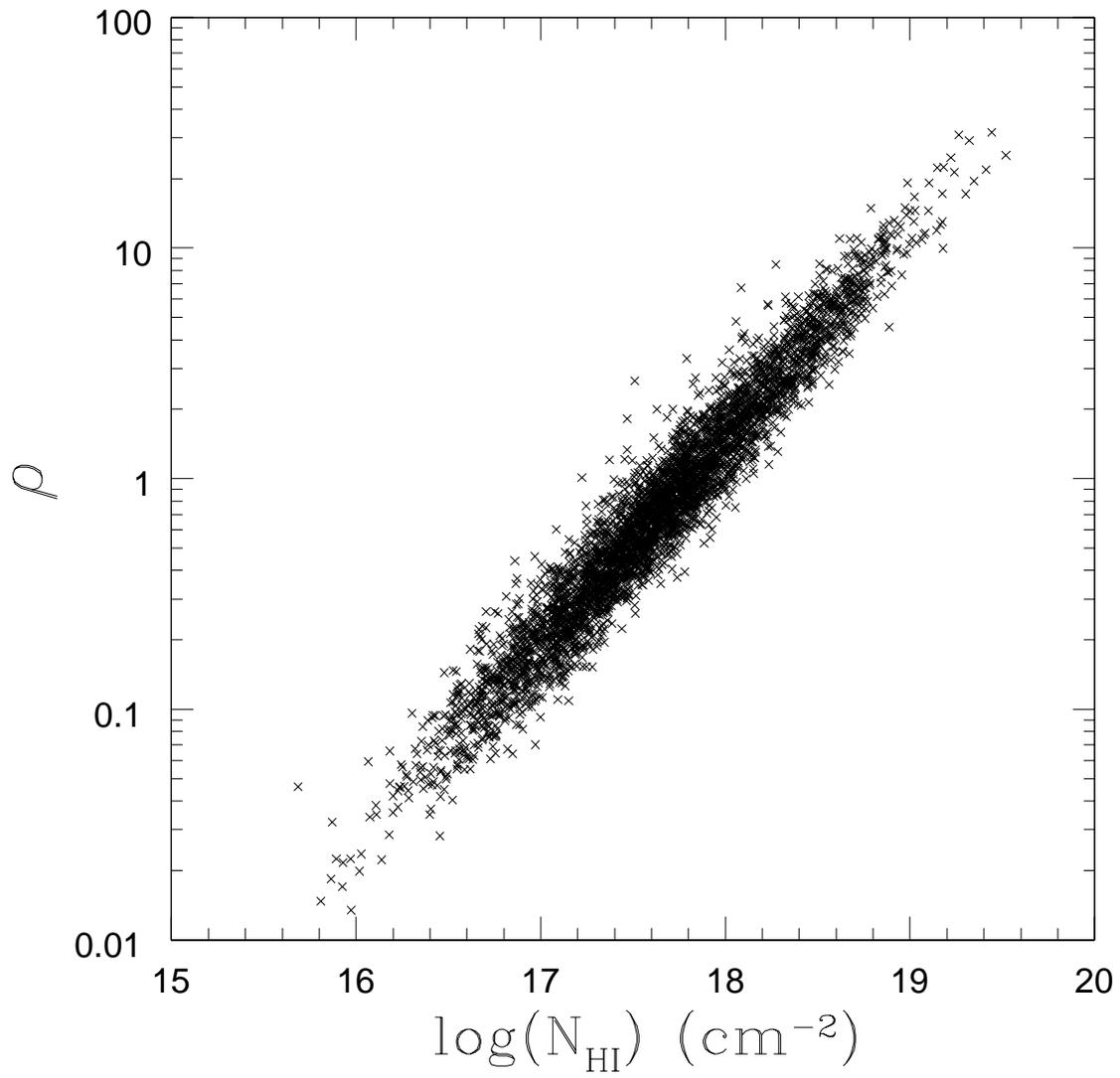
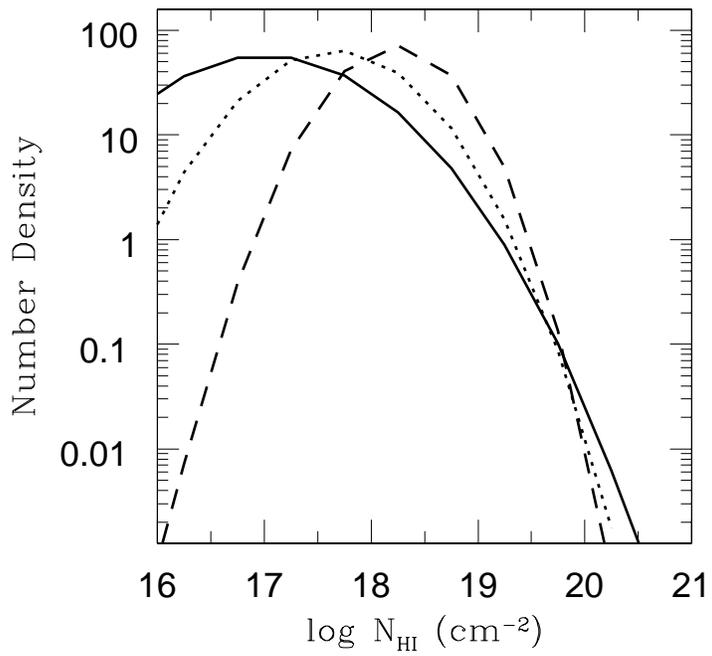
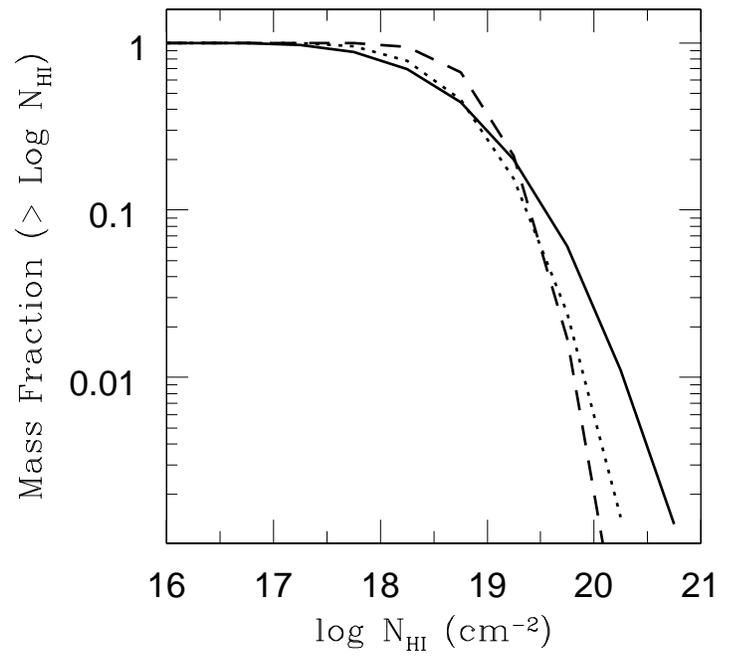
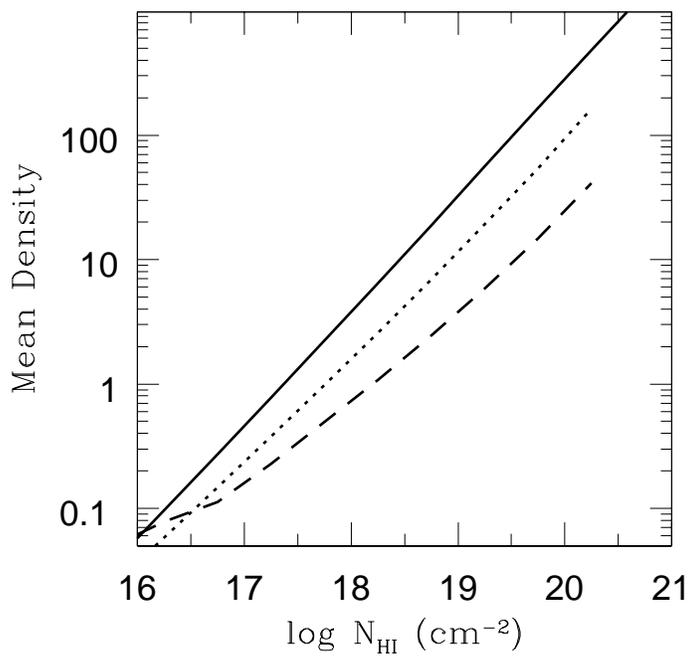
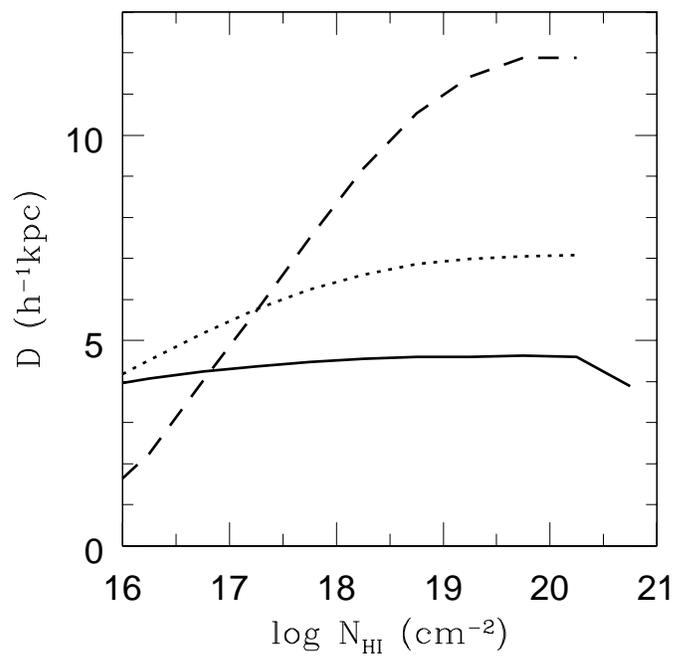
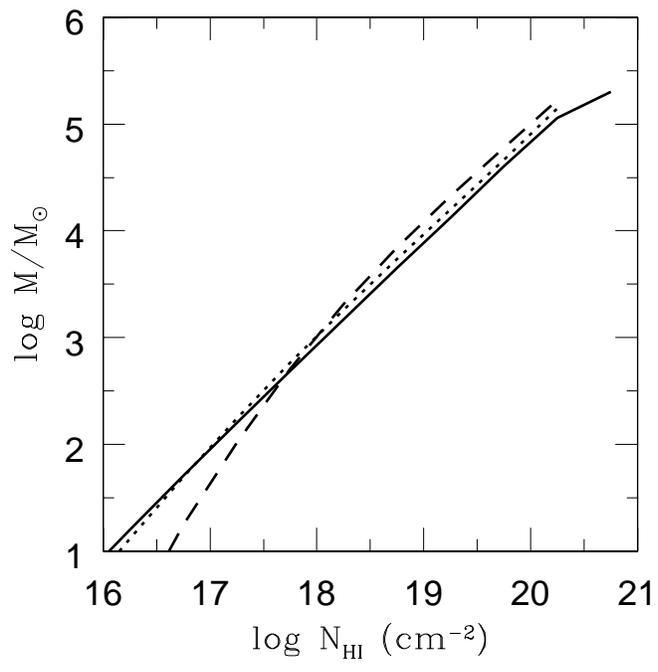
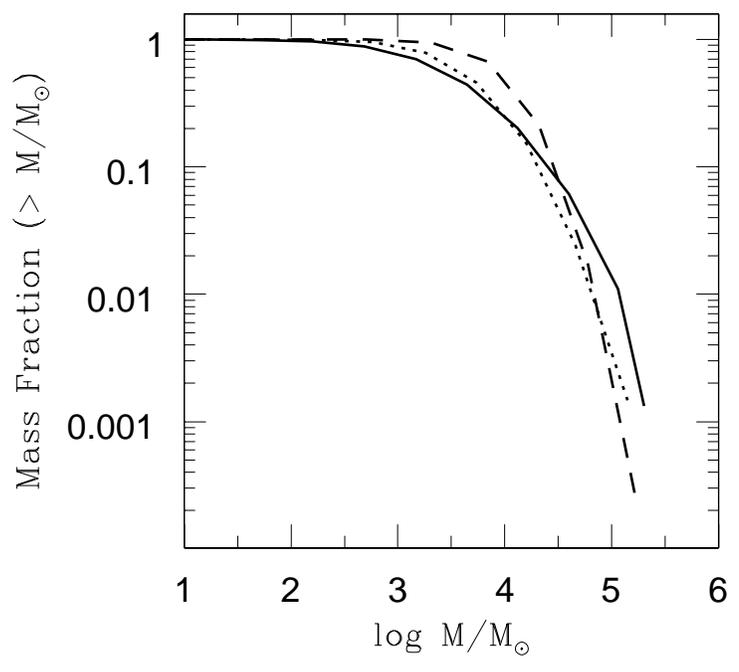
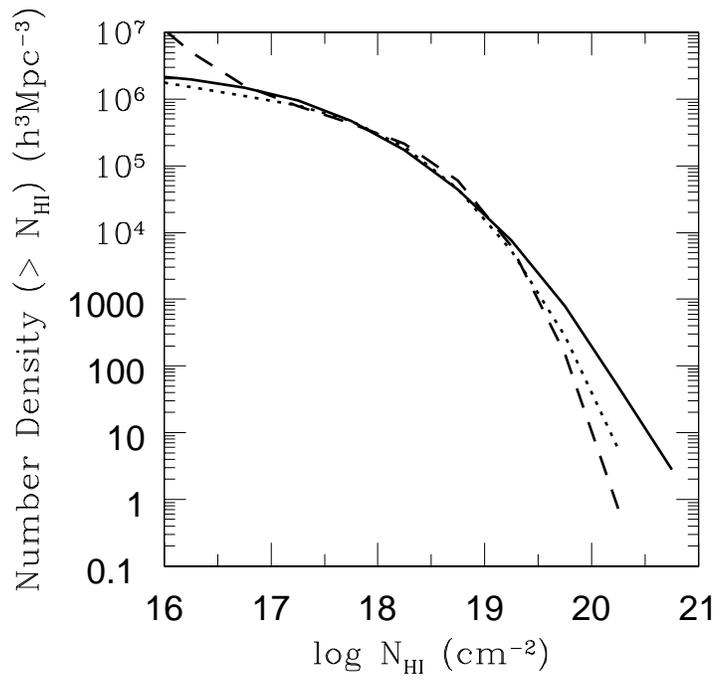
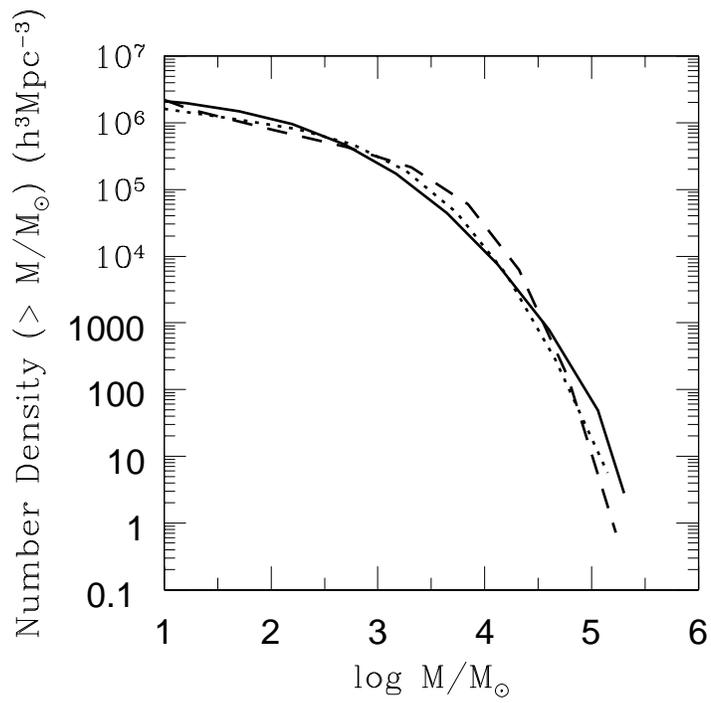
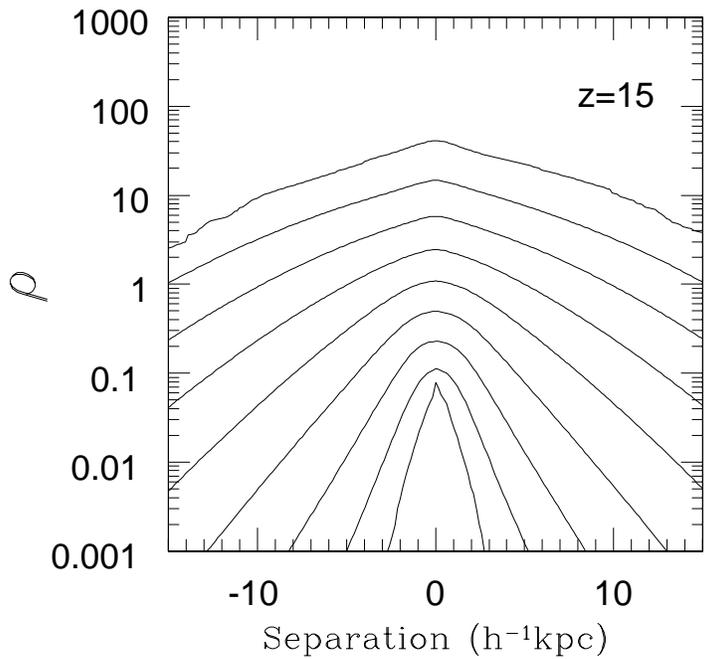
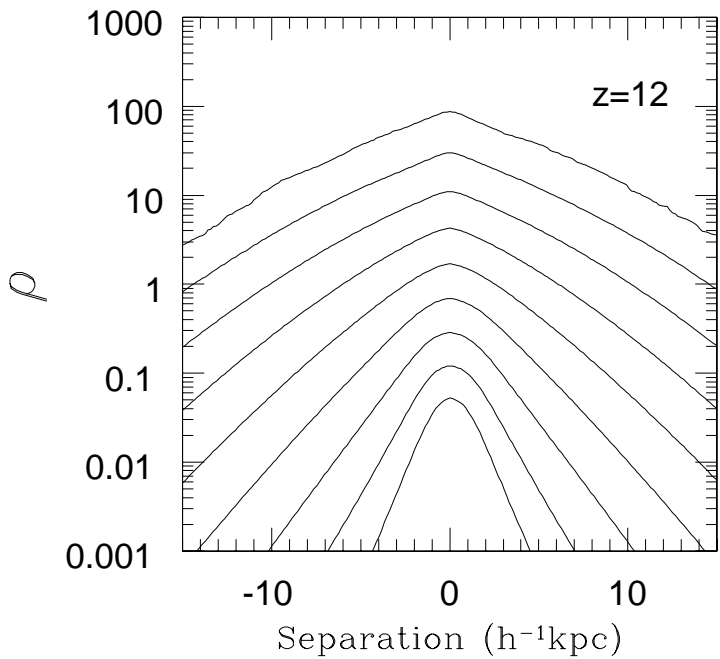
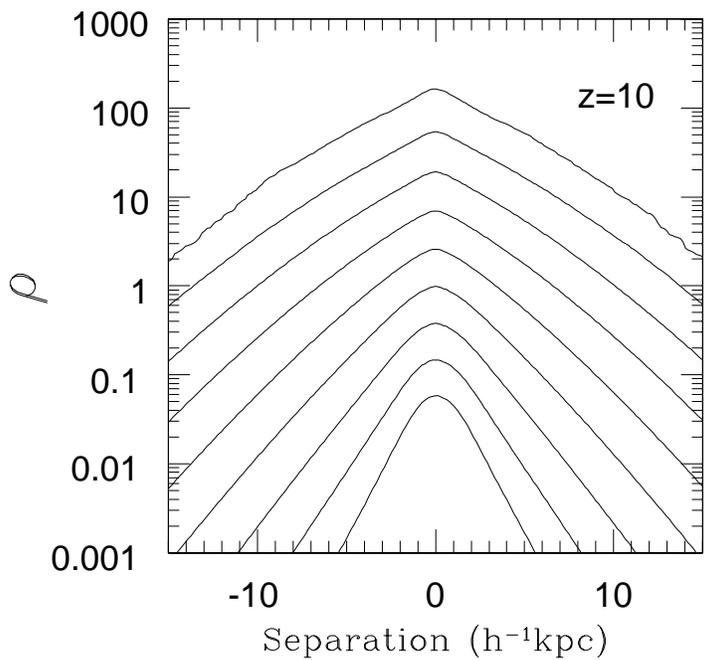
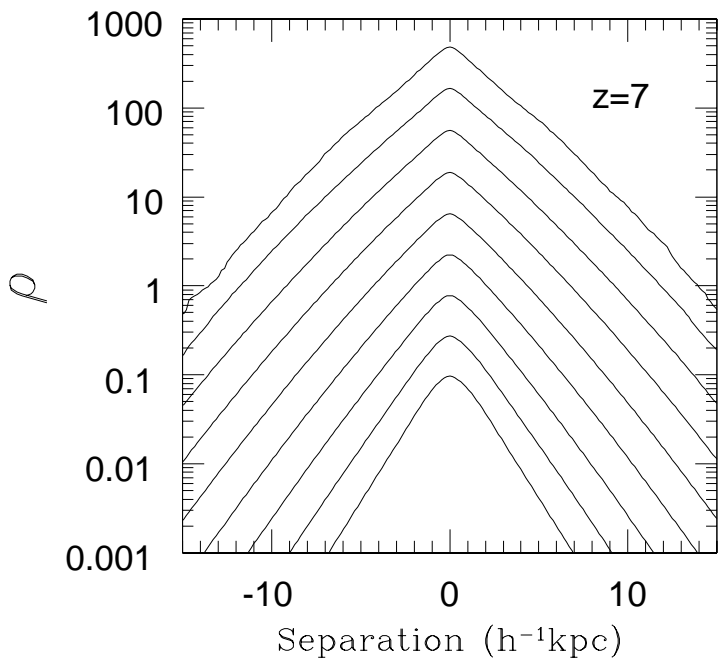
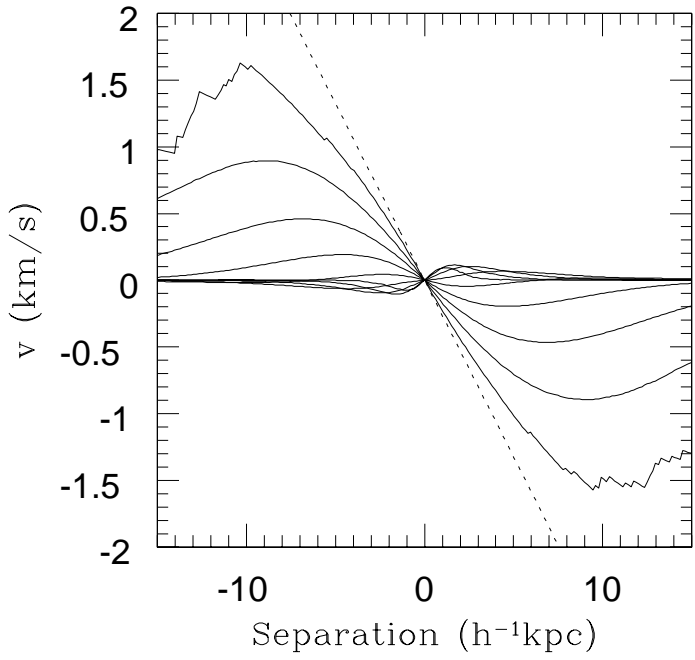
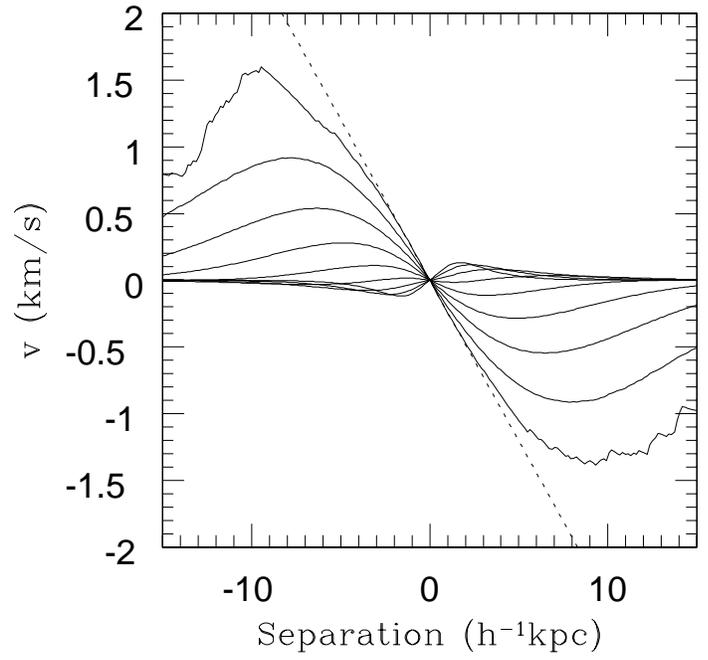
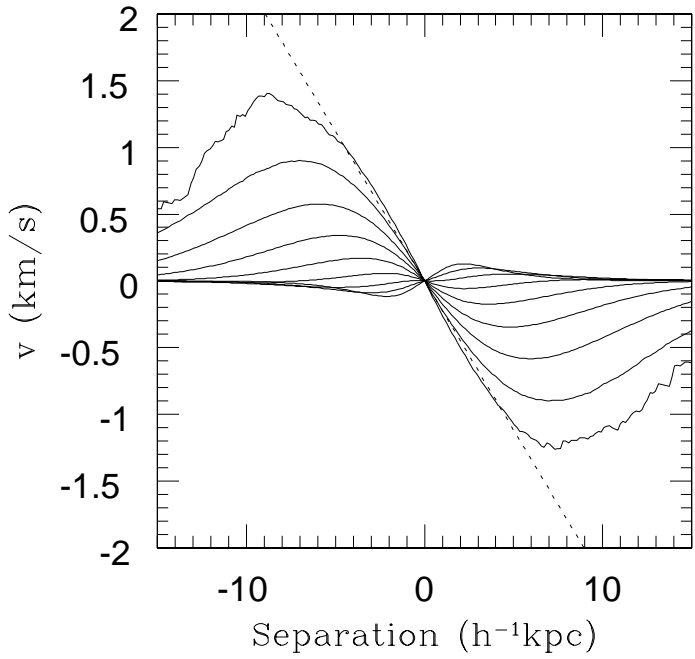
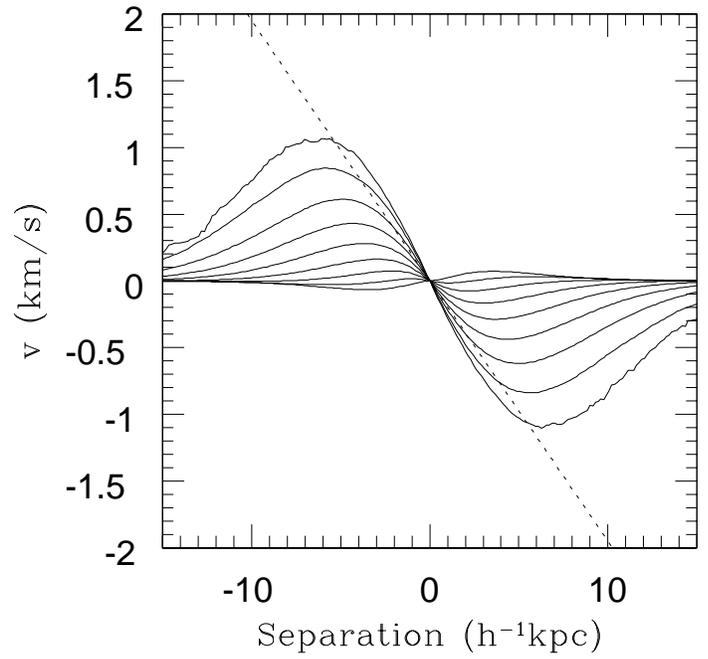
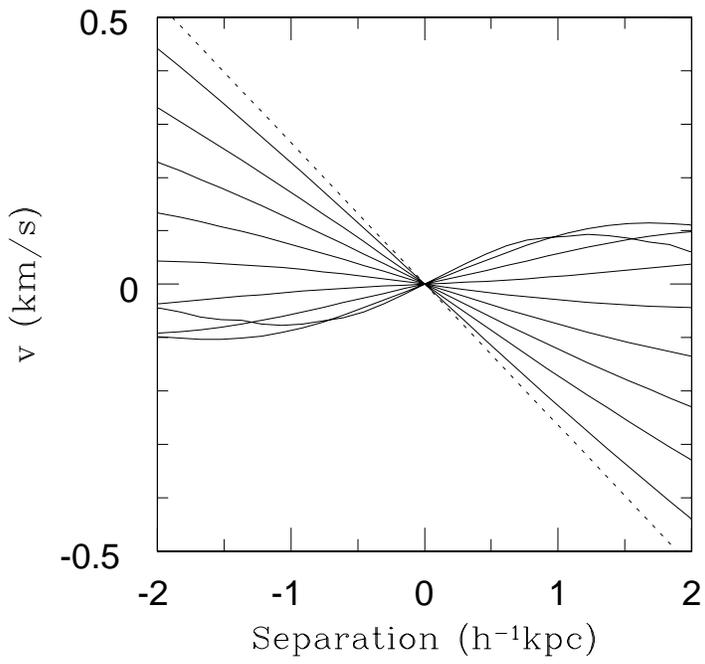
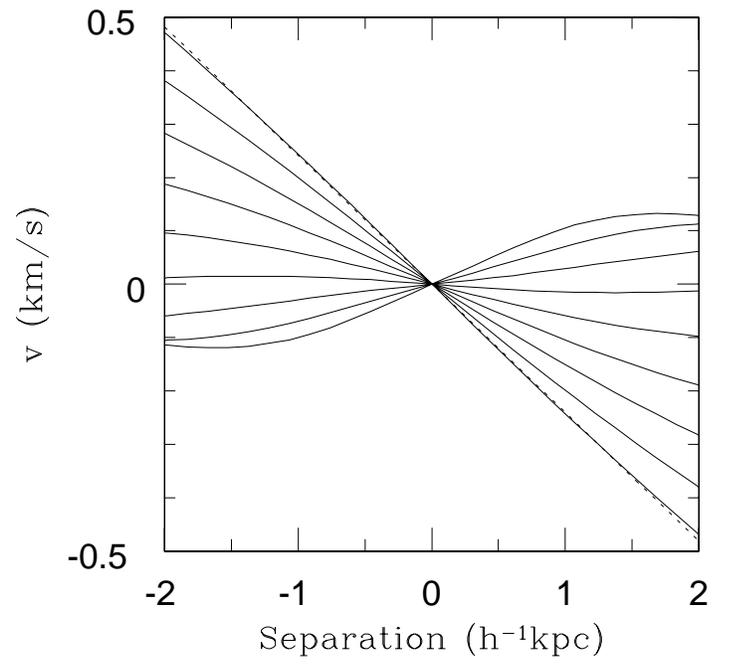
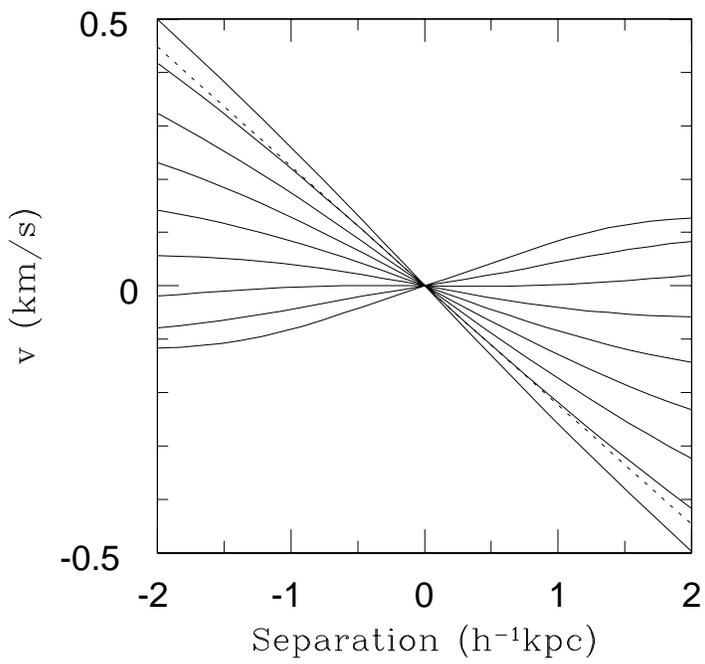
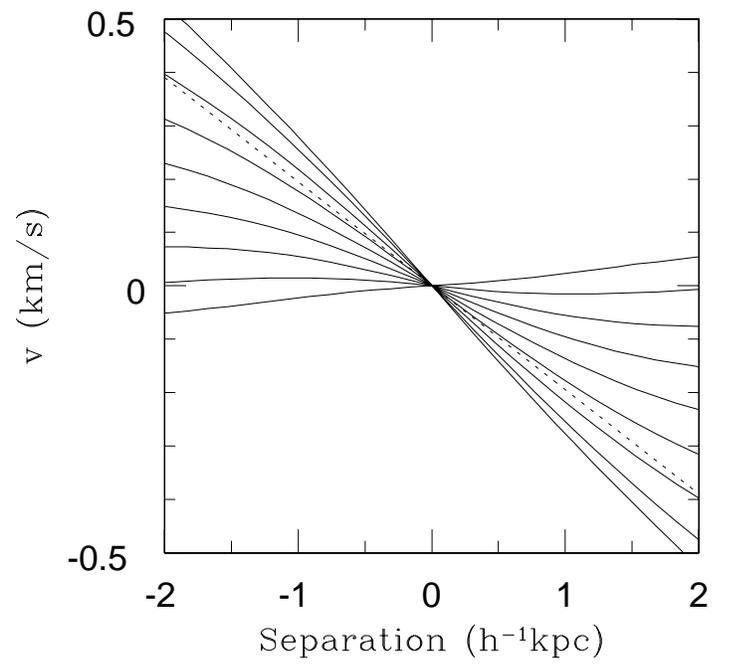
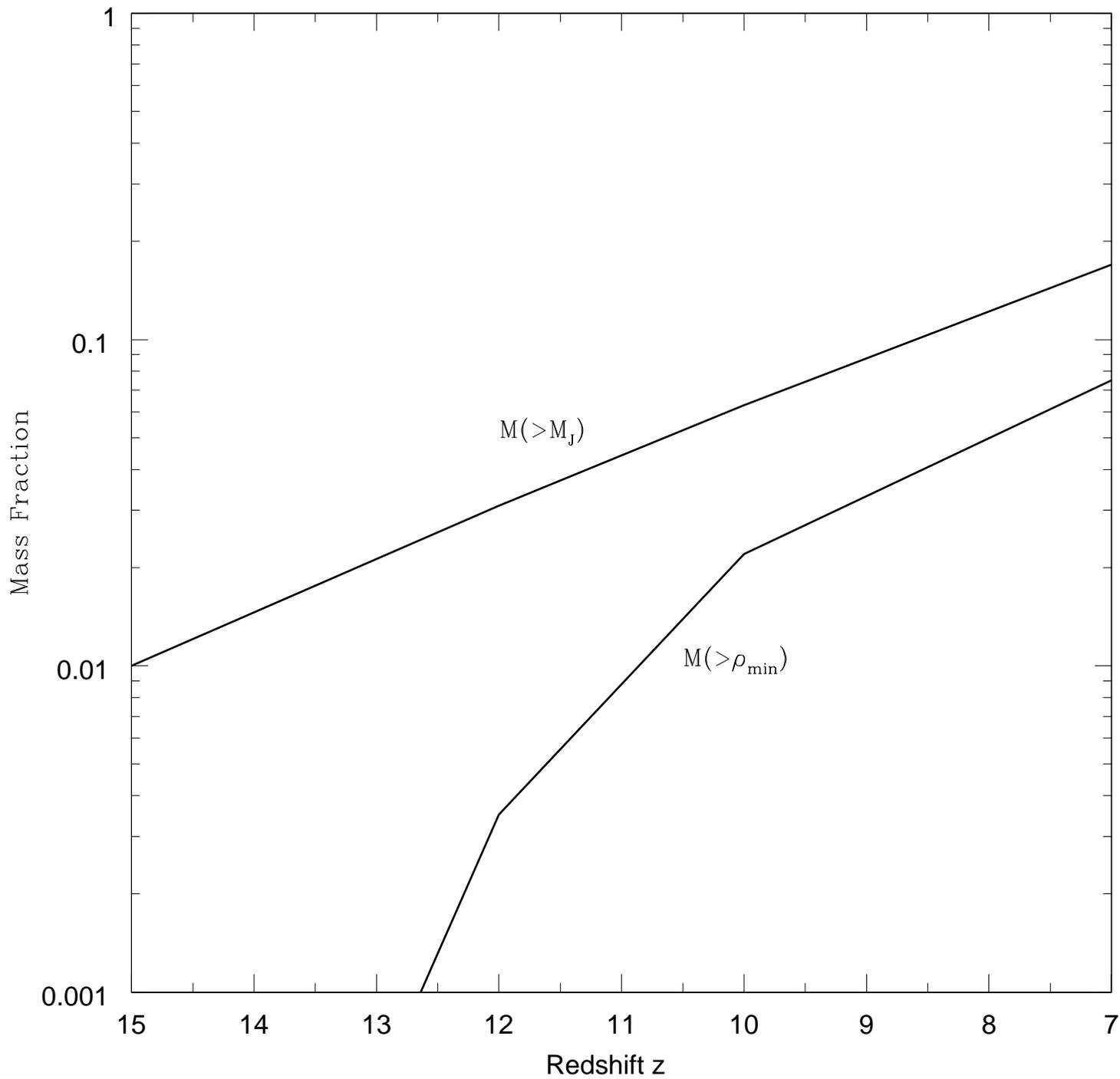
\end{document}